\documentclass[structabstract]{raa}
\usepackage{graphicx}                                   % (traditional abstract)
\usepackage{graphicx,times}
\usepackage{natbib,amssymb}
\usepackage{epsfig}
\usepackage{float}
\usepackage{setspace}%

\providecommand{\bm}[1]{\mbox{\boldmath$#1$\unboldmath}}

\begin{document}
    \title{Numerical Studies of the Kelvin-Hemholtz Instability in the Coronal Jet}
   %   \subtitle{I. Place Your Subtitle Here}
   
   \volnopage{Vol.0 (200x) No.0, 000--000}      %%preserved for Editor. DOn't remove!
   \setcounter{page}{1}          %%starting page, preserved for Editor. DOn't remove!
   
   \author{Tianle Zhao
   	\inst{1,2,3}
   	\and Lei Ni
   	\inst{1,3}
   	\and Jun Lin
   	\inst{1,3}
   	\and Udo Ziegler
   	\inst{4}
   }
   
   \institute{Yunnan Observatories, Chinese Academy of Sciences, Kunming, Yunnan 650216, China;
   	{\it leini@ynao.ac.cn}\\
   	\and
   	University of Chinese Academy of Sciences, Yuquan Road, Shijingshan Block Beijing 100049, China;\\
   	\and
   	Center for Astronomical Mega-Science, Chinese Academy of Sciences, 20A Datun Road, Chaoyang District, Beijing, 100012, China;\\
   	\and
   	Leibniz-Lnstitut f\"{u}er Astrophysik Postdam, Potsdam D-14482, Germany;\\
   }

   \date{Received~~2017 month day; accepted~~2017~~month day}
  \begin{spacing}{2.0}%%
   \abstract{The Kelvin-Hemholtz (K-H) instability in the corona EUV jet is studied via 2.5 MHD numerical simulations. The jet results from magnetic reconnection due to the interation of the new emerging magnetic field and the pre-existing magnetic field in the corona. Our results show that the Alfv\'{e}n Mach number along the jet is about 5-14 just before the instability occurs, and it is even higher than 14 at some local areas. During the K-H instability process, several vortex-like plasma blobs of high temperature and high density appear along the jet, and magnetic fields have also been rolled up and the magnetic configuration including anti-parallel magnetic fields forms, which leads to magnetic reconnection at many X-points and current sheet fragments inside the vortex-like blob. After magnetic islands appear inside the main current sheet, the total kinetic energy of the reconnection outflows decreases, and cannot support the formation of the vortex-like blob along the jet any longer, then the K-H instability eventually disappears. We also present the results about how the guide field and the flux emerging speed affect the K-H instability. We find that the strong guide field inhibits the shock formation in the reconnection upward outflow regions but helps secondary magnetic islands appear earlier in the main current sheet, and then apparently suppresses the K-H instability. As the speed of the emerging magnetic field decreases, the K-H instability appears later, the highest temperature inside the vortex blob gets lower and the vortex structure gets smaller. 
  \keywords{Sun: corona jet, K-H instability, guide-field, method: Numerical simulations 
   	}
   }
   
   \authorrunning{Tianle Zhao, Lei Ni, Jun Lin, Udo Ziegler}            %author_head in even pages
   \titlerunning{Numerical Studies of Kelvin-Hemholtz Instability in Coronal Jet}  % title_head in odd pages
   
   \maketitle
   %% The author head (on even pages) and the title head (on odd pages) will be
   %% automatically extracted from \author{} and \title{}. Whenever the title is too long,
   %% you will be asked to supply a shorter one by inserting either \authorrunning{} or
   %% \titlerunning{} before \maketitle. Anyway, you can specify your own heads.
   %%
   %%
   %% Note: In the following text body of your manuscript, please note several differences from
   %%       other major journals:
   %% (1) \subsection{Please Capitalize the First Letter of Each Notional Word in Subsection Title}
   %% (2) Please Capitalize the First Letter of Each Notional Word in all tables' captions
   
   %
   %________________________________________________ sections below
   %
   \section{Introduction}           %% first-level sections will be auto-capitalized
   \label{sect:intro}

    The jet that behaves as a transient phenomenon is ubiquitous in the solar atmosphere. It usually appears in the active region and in the polar corona hole. Jets are considered as the important mass and energy source of the upper solar atmosphere and solar wind \cite{2016SSRv..201....1R}. They are usually observed in multiple wavelengths, such as H$\alpha$, Ca II H, EUV and soft X-ray \citep[e.g.,][]{1975SoPh...42..425R,2007Sci...318.1591S,1999SoPh..190..167A,2011ApJ...735L..43S,1992PASJ...44L.173S},   and their dynamic, thermal, and structural behavious  are different when observed in different wavelengths, which can be used to explore the possible mechanisms of the jet. Coronal jets are usually observed in EUV and X-ray wavelengths. \cite{2016SSRv..201....1R} summaried the characteristics of the corona jet according to their manifestations, such as the velocity, width, height, lifetime and temperature. The temperature of EUV jets ranges from 0.1 to 10 MK, and the X-ray jets can be hotter than 10 MK. \cite{2010ApJ...720..757M} devided the jet into the €œstandard€ and  the €œblow-out€ jets according to their morphological features. The spire of the standard jet is narrow during its entire lifetime and the base is relatively dim. The standard jet are modeled by \cite{1992PASJ...44L.173S}. The spires of the blow-out jets on the other hand become broader with time and eventually reach the size comparable to the width of the jet base, the brightening of the arch base is apparent, the twist and the shearing motions usually appear in the event. Both the standard jet and the blow-out jet are considered to be resulted from  magnetic reconnection between the emerging new magnetic field and the pre-existing coronal magnetic field.  \newline
   
 As the resolution of solar telescopes are improved, many fine structures in the jet have been observed. \cite{2014A&A...567A..11Z} and \cite{2016SoPh..291..859Z} analyzed the EUV data from Solar Terrestrial Relations Observatory (STEREO) and the Atmospheric Imaging Assembly on the Solar Dynamics Observatory (SDO), they found that the bright blobs were produced and ejected out along the jets. The lifetime of these blobs is about 24$-$60 s, the temperature is between 0.5 MK and 4 MK, and the diameter ranges from  2 to 10 Mm. \cite{2017ApJ...834...79Z} recognized multiple upward and downward bright blobs in the legs of the jet via studying the high-resolution data from Interface Region Imaging Spectrograph (IRIS). Recently, several observations also show the detailed studies about the bright blobs in the corona jets (e.g. Li et al., 2017; Shen et al., 2017). The mechanism of the blob formation in coronal jets is mainly considered as the plasmoid instability  \citep[e.g.,][]{2009PhPl...16k2102B, 2013PhPl...20f1206N, 2016JPlPh..82f5901C, 2017ApJ...835..191N} in the magnetic reconnection process. In previous two-dimensional and three-dimensional simulations \citep[e.g.,][]{2012ApJ...751..152J, 2013ApJ...777...16Y, 2013ApJ...771...20M, 2016ApJ...827....4W} the high density magnetic island or magnetic flux rope (3D) were found to form in the magnetic reconnection process when the Lundquist number was high enough. These magnetic island or magnetic flux rope (3D) are believed to correspond to the observed bright blobs. 
   
 However, the magnetic island or magnetic flux rope (3D) always merged into the background magnetic fields and the plasma in these simulations, none of them was observed to be ejected out along the jet as shown in the observational results of \cite{2014A&A...567A..11Z} and \cite{ 2016SoPh..291..859Z}. For the first time,  recent high resolution numerical experiments with high Lundquist number by \cite{2017ApJ...841...27N}  indicated that the magnetic island can be easily formed and ejected out along the jet when the plasma $ \beta$ is low enough. The characteristics of the magnetic island is similar to that of bright blobs observed in the EUV band. In the case with higher plasma $ \beta$, the K-H instability appeared along the jet and resulted in the vortex-like high density and high temperature blobs, which was suggestive of accounting for the bright blob observed in the EUV band. \newline
   
 K-H instability was found on the shear plane with relative motions between two fluids by Kelvin(1871) and Helmholtz(1868) one and a half centuries ago. But only a small number of observations have shown the K-H instability on the Sun. The irregular evolution of a CME was analyzed by \cite{2013ApJ...767..170F}, they inferred that the characteristics of the evolutionary process were consistent with the result of the K-H instability.  When studying an eruption that started from an active region and produced a CME and flare, \cite{2011ApJ...734L..11O} noticed for the first time that the K-H instability occuring in the eruption magnetic configuration. They found a set of vortices at the interface between the region where the eruption took place and the nearby region.  The sizes of these vortices varied from a few to ten arcseconds and the speed of these features on the interface ranged from 6 to 14 km/s. \cite{2011ApJ...734L..11O} identified the vortices with the consequence of the K-H instability, which was confirmed by their linear analysis and nonlinear 2.5-MHD numerical experiment.  They concluded that it is the velocity shearing between the erupting and the nearby stationary magnetic configuration that drove the instability. \newline
 
 \cite{2016ApJ...830..133K} found that some of the small-scale structures in the chromospheric jet displayed apparent red and blue shift in the spectral lines and that the vortex-like structures rapidly appeared along the boundary of the jet. On the basis of the spectral analysis, they found that the chromospheric spectral lines became broader inside these vortex-like and turbulent structures, which could be ascribed to the K-H instability. \cite{2015ApJ...813..123Z} compared the results of their theoretical model with those of observations, and found that when the jet speed was higher than the Alfv\'{e}n speed in the jet direction, the K-H instability may happen. So far, both theory and observations indicate that the K-H instability is not easy to happen in the X-ray jets with high speed and low plasma density, but it is easier to appear in the EUV jets with lower speed and higher plasma density. \newline
   
  Some studies about the  basic theories and numerical simulations of the K-H instability in the case with magnetic fields along the shearing flows have been reported. In the simulations by \cite{1997ApJ...482..230J} and \cite{2000ApJ...529..536J}, the K-H vortices were found to persist until the viscosity and small-scale reconnection dissipated them when the plasma $\beta$ was high enough ($\beta=3000$ \& $\beta=24000$) and the magnetic field was weak. \cite{1999JPlPh..61....1K} found that in a uniform magnetic field, the K-H mode grew between two shearing flows with time when the Alfv\'{e}n Mach number $M_A=9$, but it was stabilized when  $M_A=1.5$. \newline
   
   \cite{2016ApJ...824...60T} numerically explored the K-H instability in the case of different magnetic fields imposed in the direction of the two shearing flows. The results showed that the dynamic behaviours of the plasma fluid change with the Alfv\'{e}n Mach number in the direction of the fluid velocity. The K-H mode was linearly stabilized for $M_{A}\leq2.27$; the K-H mode was nonlinearly stable and developed into wavy motions for $2.27\leq M_{A} \leq 2.8$; in the range of $2.8 \leq M_{A} \leq 6.2$, the K-H mode was unstable and evolved into filamentary flows; the K-H vortex can fully roll up in an even higher $M_{A}$, e.g. $M_{A}=50$, but the small-scale reconnection would destroy the K-H instability soon. The K-H instability has not
been identified in the previous corona EUV jet simulations except for the work by \cite{2017ApJ...841...27N}.\newline
   
 In this work, we study the K-H instability in the solar coronal jet with different guide fields based on the previous 2D work by \cite{2017ApJ...841...27N}. The effects of guide-field and flux emerging speed on the jet formation and K-H instability process will be presented. The numerical model is described in Section 2. We will present our numerical results in Section 3. In the last section, we will summary this work.\newline

   %% Authors can give a citation as 'Michel et al. 1992'.
   %% You may also use \cite, \citep and \citet for citation, and use Table~1 or Figure~1
   %% and so forth. Using \ref and \label for cross-references of Tables/Figures
   %% is a good way in adjusting/adding/removing text, tables or figures.

   \section{Numerical Model}
   \label{sect:Obs}
   
   The single-fluid MHD equations including the gravity and the thermal conduction are given below:\newline
  \begin{eqnarray}
 \partial_t \rho &=& -\nabla \cdot \left(\rho \mathbf{v}\right),                                                                \\ 
 \partial_t \mathbf{B} &=& \nabla \times \left(\mathbf{v} \times \mathbf{B}-\eta\nabla \times \mathbf{B}\right), \label{e:induction}\\
  \partial_t (\rho \mathbf{v}) &=& -\nabla \cdot \left[\rho \mathbf{v}\mathbf{v}
                              +\left(p+\frac {1}{2\mu_0} \vert \mathbf{B} \vert^2\right)\mbox{\bfseries\sffamily I} \right]  \nonumber \\
                              & &+\nabla \cdot \left[\frac{1}{\mu_0} \mathbf{B} \mathbf{B} \right] + \rho \mathbf{g}, \\
 \partial_t e &=& - \nabla \cdot \left[ \left(e+p+\frac {1}{2\mu_0 }\vert \mathbf{B} \vert^2\right)\mathbf{v} \right] \nonumber\\
       & &+\nabla \cdot \left[\frac {1}{\mu_0} \left(\mathbf{v} \cdot \mathbf{B}\right)\mathbf{B}\right] + \nabla \cdot \left[ \frac{\eta}{\mu_0} \textbf {B} \times \left(\nabla \times \mathbf{B}\right) \right] \nonumber\\
       & & -\nabla \cdot \mathbf{F}_\mathrm{C}+\rho \mathbf{g} \cdot \mathbf{v},   \\
   e &=& \frac{p}{\Gamma_0-1}+\frac{1}{2}\rho \vert  \mathbf{v} \vert^2+\frac{1}{2\mu_0}\vert \mathbf{B} \vert^2,          \\
   p &=& \frac{2\rho}{m_\mathrm{i}} k_\mathrm{B}T .
\end{eqnarray}
Here, $\rho$, $\mathbf{v}$, $e$, $\mathbf{B}$ and $p$  represent the plasma density, velocity, the total energy density, the magnetic field and  the gas pressure respectively. $\mathbf{F}_\mathrm{C}$ is the flux of the thermal conduction. $\mathbf{g}=-273.9~\mbox{m\,s}^{-2}~\mathbf{e}_y$ is the constant acceleration of gravity of the Sun. In this work, we use the international system of units (SI) for all the variables.\newline
   
   The initial background magnetic field is set as $B_{x0}=-0.6b_0$ and $B_{y0}=-0.8b_0$ ($b_0=0.0015$~T). In this work, we have added different guide fields in the ${z}$-direction for four cases: $B_{z0}=€–0.05b€—_0$ in case I, III and IV and $B_{z0}=b_0$ in Case II. The initial plasma velocity is zero, and the initial temperature is $T_0=8\times10€—^5$~K. The initial stratified density including the constant acceleration of gravity is given as   
\begin{equation}
  \rho_0=\rho_{00}\mathrm{exp}\left(-\frac{{m_i} {g}}{2{k_B}{T_0}}{y}\right)
\end{equation} 
 where $\rho_{00}= 0.5\times1.66057\times10^{-10}$~kg\,m$^{-3}$, the mass of proton is $m_\mathrm{i}=1.66057\times10^{-27}$~kg and the Boltzmann constant is $k_\mathrm{B}=1.3806\times10^{-23}$~J\,K$^{-1}$. The simulation box is inside the domain $ 0\textless x\textless200L_0 $ and  $0\textless y\textless100L_0$, with $L_0=10€—^6$~m.\newline
   
  We use the temperature-dependent magnetic diffusivity in all the four cases:
\begin{equation}
   \eta=10^8\left(T_0/T\right)^{3/2}+10^9\left[1-\mathrm{tanh}\left(\frac{y-2L_0}{0.2L_0}\right)\right].
 \end{equation}
Unit of $\eta$ is ~m$^2$\,s$^{-1}$. The anisotropic heat conduction flux, $\mathbf{F}_\mathrm{C}$, is given by (e.g., see also Spitzer 1962):
\begin{eqnarray}
\mathbf{F}_\mathrm{C}=-\kappa_{\parallel}\left(\nabla T\cdot \hat{\mathbf{B}}\right)\hat{\mathbf{B}}-\kappa_{\perp}\left[\nabla T-\left(\nabla T\cdot \hat{\mathbf{B}}\right)\hat{\mathbf{B}}\right],
\end{eqnarray}
where $\hat{\mathbf{B}}=\mathbf{B}/ \vert \mathbf{B} \vert $ is the unit vector in the direction of the magnetic field. The parallel and perpendicular thermal conductivity ratio, 
$\kappa_{\parallel}$ and $\kappa_{\perp}$, are given by:
\begin{eqnarray}
\kappa_{\parallel}&=&\frac{1.84\times10^{-10}}{\ln \Lambda} {T}^{5/2},\\
\kappa_{\perp} &=& \min (\kappa'_{\perp}, \kappa_{\parallel}),
\end{eqnarray}
with
\begin{eqnarray}
\kappa'_{\perp} = 8.04\times10^{-33} \left(\frac{\ln \Lambda}{m_\mathrm{i}}\right)^2 \frac{\rho^2}{T^{3} B^{2} } \kappa_{\parallel}, \nonumber
\end{eqnarray}
where $\ln \Lambda=30$, unit for $\kappa_{\parallel}$ and $\kappa_{\perp}$ are J~K$^{-1}$~m$^{-1}$~s$^{-1}$.
  \newline
   
   We use the NIRVANA code to solve equations (1) through (11) in this work. This code has been clearly described in previous works  \citep[e.g.,][]{2008CoPhC.179..227Z, 2011JCoPh.230.1035Z, 2017ApJ...841...27N}. The adaptive mesh refinement method in this simulation is the same as in the paper by Ni et al. (2017), the base-level grid is $320\times160$, the highest refinement level is 10. All the pictures presented in this work are also plotted by using the level 3 or level 4 uniform IDL data, which are transformed from the original raw data. \newline
  
  Two extra layers with the ghost grid cell are applied to the code to set boundary conditions at each boundary. The boundary conditions are the same as in the previous paper by \cite{2017ApJ...841...27N} except that we also need to set boundary conditions for magnetic field and velocity in $z$-direction in this work. The outflow boundary conditions  as described in \cite{2017ApJ...841...27N}  are applied at the left ($x=0$) , right ($x=200L_0$) and up ($y=100L_0$) boundaries. The condition of divergence-free of the magnetic field requires the continuity of the normal component of the magnetic field on the boundary, which can be used to extrapolate the normal component through the boundary. We also insert two ghost layers below the physical bottom boundary $y=0$. The gradient of the plasma velocity vanishes at the bottom boundary. The magnetic field inside the two layers with the ghost grid cells are set as:

\begin{equation}
	b_{xb}=-0.6b_0+\frac{100L_0(y-y_0)b_1 f}{[(x-x_0)^2+(y-y_0)^2]}\left[\mathrm{tanh}\left(\frac{x-70L_0}{\lambda}\right)-\mathrm{tanh}\left(\frac{x-130L_0}{\lambda}\right)\right]    , 
	\end{equation}
	\begin{equation}
	b_{yb}=-0.8b_0-\frac{100L_0(x-x_0)b_1 f}{[(x-x_0)^2+(y-y_0)^2]}\left[\mathrm{tanh}\left(\frac{x-70L_0}{\lambda}\right)-\mathrm{tanh}\left(\frac{x-130L_0}{\lambda}\right)\right]    , 
	\end{equation}

\noindent where $f=t/t_{1}$ for $t\le t_{1}$ and  $ f=1 $ for $ t\ge t_{1}$, $ x_{0} = 100L_{0} $, $y_{0}=-12L_{0} $, $b_{1}=3\times10^{-4}$~T and $\lambda=0.5L_{0}$ with $t_{1}=500$~s for Cases I and II, $t_{1}=700$~s for Case III, and $t_{1}=350$~s for Case IV, respectively. We set the magnetic field in the $z$-direction in the bottom ghost grid cells $b_{zb}$ to equal to the initial value of the magnetic field in the $z$-direction inside the compulational domain, $b_{zb}=B_{z0}$. When $t<t_{1}$, the strength of the magnetic field below the bottom boundary varies with time, flux emerging stops after $t=t_{1}$. As shown previously by \cite{1984SoPh...94..315F}, \cite{2000ApJ...545..524C}, and \cite{2010A&A...510A.111D}, one can set up the magnetic flux emergence by changing the conditions with time at the bottom boundary. As described by \cite{2017ApJ...841...27N}, the magnetic field does not fulfill the divergence free condition $\nabla \cdot \mathbf{B} = 0$ inside the two ghost layers around $x=70L_0$ and $x=130L_0$. The high magnetic diffusion below $y=0.2L_0$ as shown in equation (8) can smooth the non-physical features inside the two ghost layers, then the related values at the bottom boundary are smoothed.\newline

  \section{Numerical Results}
   \label{sect:data}
   \subsection {K-H instability in corona jet with guide field}
   
 Case I and Case II differ from each other in the guide field. The guide field is $B_{z0}=0.05b_0$ for Case I, and  $B_{z0}=b_0$ for Case II. In Figures~1 and 2 we can see the distribution of the current density in $z$-direction $J_z$, the temperature $T$, the plasma density $n$, and the velocity along the jet direction $v_{||}$ and the distributions of the emission count rate in the AIA 211\,\AA\, channel at five different times in Cases~I and II, from which we can see the jet evolutionary process in the two cases after the emergence of magnetic field stops. The distributions of each variable in Case~I as shown in Figure 1 are almost the same as those displayed in Figure 7 of  \cite{2017ApJ...841...27N}. The jet lifetime is about 37 minutes, the jet maximum temperature  is about 1.8 MK, and the maximum velocity along the jet direction is 320~km~s${^{-1}}$. \newline
 
 In Case II, the jet lifetime is about 33 mins, the maximum temperature is 1.6 MK, and the maximum velocity along the direction is 275~km~s$\mathrm{^{-1}}$, which are smaller than the corresponding parameters in Case I. In Figure~3, we presented the maximum velocities along the jet direction at each time step in all the four cases. One can see that there is an apparent peak in Cases I, III and IV with weak guide field, these peaks appear after the K-H instability starts and before the magnetic island appears in each of these case. However, there is no apparent peak in Case II with strong guide field. The maximum velocity along the jet direction in Case II is also slightly smaller than that in Case I after K-H instability appears. We notice that the characteristics of the simulated jets here are consistent with those showed by EUV observations of  corona jets (see Raouafi et al. 2016).  \newline
   
In Figures 1 and 2, we can see clearly that the vortex-like blobs with high density and high temperature  are rolled up in the jet. The vortex-like structure of the plasma blob at the bottom of the jet is more apparent than those at the higher positions of the jet. The higher the plasma blob is, the less the blob is rolled up. As pointed out by \cite{2017ApJ...841...27N}, these vortex-like blobs indicate the strong shearing flows between the the surrounding plasma and jet, which leads to the K-H instability.   Based on the previous theory and simulations \citep[e.g.,][]{1999JPlPh..61....1K, 2016ApJ...824...60T}, the K-H instability can be suppressed by the magnetic field along the shearing layers, and it can only appear when the Alfv\'{e}n Mach number along the shearing layers is high enough. Figure~4 shows the distribution of velocity $v_{||}$ and the corresponding Alfv\'{e}n Mach number, $M_A$, at $t=784.3$~s before K-H instability happens. $v_{||}$ is the plasma velocity along the jet direction. From Figure~4, we can find that the jet direction is roughly the same as the direction of the initial background magnetic fields in the $xy$- plane at $t=784.3$~s. Though the maximum of $v_{||}$ is only 265~km~s${^{-1}}$, which is still larger than the corresponding Alfv\'{e}n velocity $v_{A||}$ along the jet. The Alfv\'{e}n Mach number along the jet direction is defined as $M_A=v_{||}/v_{A||}$ \citep[e.g.,][]{1999JPlPh..61....1K, 2016ApJ...824...60T}, and the value of $M_A$ is between 5 and 14 in most regions inside the jet. In this work, we use  the strength of magnetic field component which is parallel to the jet direction to calculate the Alfv\'{e}n speed $v_{A||}$. From Figure 4 we can also find that $M_A$ is apparently larger than 14 at some local areas, the magnetic fields are strongly folded and almost perpendicular to the jet direction at these areas. Therefore, the Alfv\'{e}n velocity $ v_{A||}$ along the jet is close to zero and $M_A$ is very large in these areas. \newline
  
 \cite{1999JPlPh..61....1K} studied the K-H instability by using different initial magnetic configurations around the shearing flows. In the cases that parallel magnetic fields at both sides of the flows possess the same orientation, they found that the K-H instability grew with time when the initial Alfv\'{e}n Mach number was $M_A=9$, and it was stabilized when the magnetic field was strong and $M_A=1.5$. \cite{2016ApJ...824...60T} concluded that the K-H mode was unstable and evolved into filamentary flows when $2.8 \leq M_{A} \leq 6.2$, but  the K-H vortex can fully roll up only for a very large value of $M_{A}$, say $M_{A}=50$. In both of the two papers, $M_{A}$ was calculated by using the initial flow velocity and Alfv\'{e}n velocity along the direction of shearing flows. \newline

Comparing our analysis in the previous paragraphs with their results, we find that the value of $M_A$ in the jet region in our simulations are always big enough to trigger the K-H instability. But this value is not big enough in most areas in the case as shown in Figure~4 to fully roll-up the plasmas, especially in the regions near the top of the jet. \cite{2016ApJ...824...60T} also found that the small scale reconnections between the rolled-up magnetic fields would destroy vortex-like structures soon after the vortex structures were formed, even for a large $M_A$. We notice as well many small scale current sheet  fragments due to magnetic reconnection inside the vortex-like blobs as shown in Figures 1 and 2, which also play a role in destroying the vortex blob. The appearence of these vortex-blobs last for about 8 minutes in Case~I and about 6 minutes in Case II. \newline

 \cite{2015ApJ...813..123Z} analyed and derived the critical conditions for the K-H instability in both the axial and the azimuthal directions of the jets.  On the basis of the observation results and their theoretical models, they studied the possibilities for generating the K-H instabilities in macro-spicules, type II spicules, X-ray coronal jets and EUV jets. They concluded that the K-H instability is more likely to appear in the higher density and lower speed EUV jet , the K-H instability can still occur along the EUV jet with plasma density up to $10^{16}-10^{17}$~m$^{-3}$ and  velocity of 250~km~s$^{-1}$ when the axial magnetic field is about 10~G and the Alfv\'{e}n speed reaches 220~km~s${^{-1}}$. For comparision, we listed several important parameters with the corresponding characteristic values obtained in this work and by Zaqarashvili et al (2015) in Table 2. We notice that the result deduced by two works are  consistant with one another.\newline
   
   From Figures~1, 2, 5 and 7, we can see that the reconnection outflow in the main current sheet and the vortex-like structures are very different from one another in Cases I and II. The plasma velocity divergence $\nabla \cdot \mathbf{v}$ reflects the compression degree  of the plasma. From the distributions of $\nabla \cdot \mathbf{v}$ in the simulation domain, we can preliminarily judge if the shock structure appears or not \citep{2017ApJ...841...27N, 2016ApJ...822...18N}. We have analyzed the distributions of $\nabla \cdot \mathbf{v}$ at different times in Cases I and II. In addition to the intermediate shock at shock front SF1 as shown in Figure~5, we have recognized two fast mode shocks at shock front SF2 and shock front SF3 in the outflow region of the main current sheet in Case~I with weak guide field. We only find the intermediate shock at shock front SF1 in the whole evolution process of Case~II with strong guide field, and no fast mode shock appears. \newline
   
   We have used the MHD jump conditions as presented below (e.g., see also \cite{2017ApJ...841...27N}) to analyze these shocks and judge the type of shocks. In MHD jump conditions (e.g., see also Priest 2014):  \begin{eqnarray}B_{n1}&=&B_{n2},\\ \rho_1 v_{n1}&=&\rho_2 v_{n2},\\ \rho_1v_{n1}^2+p_1+\frac{B_{t1}^2}{2\mu_0}&=&\rho_2v_{n2}^2+p_2+\frac{B_{t2}^2}{2\mu_0},\end{eqnarray} where subscript $t$ represents the component which is tangential to the shock front, subscript $n$ represents the component that is normal to the shock front, properties ahead of and behind the shock are denoted by 1 and 2. From Figures~5(a) and 5(b), we can see that the plasma is sharply heated to high temperatures behind the two fast mode shocks in Case I. But no fast mode shock occurs in Case II. However, heating that was probably caused by the compression and Joule disspation could also be noticed (Figures 5(c) and 5(d)).\newline
    
 In both Cases~I and II, the vortex-like structures start to break after the magnetic islands have appeared in the main current sheet as shown in Figures~1 and 2. Figure~6 shows how the vortex-like structures change before and after magnetic islands appears. One can find that the vortex-like blobs start to move toward the left bottom and break  after the magnetic islands appear.  Figure~6(b) shows the distribution of the kinetic energy along the jet direction at different times. As  multiple reconnection X-points and magnetic islands appear in the main current sheet, the upward reconnection outflow velocity and the corresponding kinetic energy gradually decrease. When the  outflows cannot provide enough kinetic energy to push the high density vortex-like blobs upward along the jet, these vortex-like blobs start to fall down. So, in addition to small scale magnetic reconnection inside the vortex like blob, magnetic islands appearing in the main current sheet is another important reason to cause the K-H instability to disappear.  \newline
 
   As shown in Figure 6(b), the thick lines NL1 and NL2 are located along the current sheet direction at $t=1041.3$~s before magnetic islands appear and after magnetic islands appear at $t=1294$~s, respectively. Figures~6(c) and 6(d)  show the distributions of the plasma velocity along the NL1 and NL2, they show that the outflow velocity decreases apparently after magnetic islands appear. In the turbulant reconnection process many magnetic islands of different sizes, together with the X-points, occur inside the current sheet  (e.g., see also Bhattacharjee et al., 2009; B\'{a}tar et al., 2011; Shen et al., 2011; Mei et al., 2012; Ni et al., 2015).  The reconnection outflow  bifurcates at the X-point, which makes the different sizes of magnetic islands connecting to the nearby X-point  have different velocities (some of them even have  opposite  velocities). Therefore, these islands  collide and coalesce with one another, and their motions spontaneously slow down. This issue has also been discussed previously \citep[e.g., see][]{2015ApJ...813...86I}.  \newline

 As shown in Figure~7, three vortex-like blobs could be recognized clearly in Case~I, but only two obvious vortex-like blobs are identified in Case~II. Comparing Figure~1 with Figure~2, we notice that the magnetic island does not appear yet in the main current sheet at $t=1186.7$~s in Case~I, but several magnetic islands already appear in the main current sheet at the same time in Case~II. Magnetic islands in the main current sheet of Case~II appear about 3~min earlier than those in Case~I. Therefore, the third vortex-like blob that was supposed to appear at the top of the jet in Case~II did not show up before the two vortex-like blobs at the lower position had already been destroyed.\newline
 
  In order to compare with observations and  previous results \citep{2017ApJ...841...27N} , we calculate the temperature and density dependent emission count rate, the emission count rate is calculated as $\mathrm{ECR}=\int n^2f(T)dl$~DN\,s$^{-1}$\,pixel$^{-1}$, where $f(T)$ is the AIA 211\,\AA\ response function, $n$ is the number density, and $dl$ is the line element along the line of sight  (e.g. see also Ni et al. 2017).  From the AIA synthetic EUV images shown in Figure~1(e), we can see that there are three obvious vortex-like bright blobs at $t=1044.8$~s and $t=1186.7$~s in Case I with weak guide field, which are similar to those shown by \cite{2017ApJ...841...27N}. As shown in Figure~2(e), the vortex-like blobs are less obvious and we can only identify two bright blobs caused by the K-H instability in Case II with strong guide field. \newline

 Ni et al. (2013) investigated the impact of guide field on the reconnection process. They showed that different guide fields result in different critical values of the Lundquist number. Magnetic islands can only appear when the Lundquist number  exceeds such a critical value. Including  guide field in the reconnection region changes the distributions of the plasma  and the magnetic pressures. The results of this work indicate that the main current sheet reaches the critical Lundquist number earlier in the case with strong guide field. The shock structures obviously appear in the upward outflow regions of the main current sheet in Case I with weak guide field, but no apparent shock structure is found in Case II with strong guide field. It is the influence of guide field on the reconnection process as discussed above that causes the K-H instability and the vortex-like blobs to form in different fashions. \newline

\subsection{Inpact of the flux emerging speed to the jet and the K-H instability}
  The important parameters in this work are listed in Table~1. The guide field used in Cases I, III, and IV is the same, but the emerging times in these cases were different. The total emerging time in Case III is 700~s,  in Case I is 500~s and it is only 350~s in Case~IV. Therefore, the flux emerging in Case~IV is faster than that in Case~I, and the flux emerging in Case~III is the slowest one. Figure 8 shows the variations of the electromagnetic energy emerging through the bottom boundary versus time in Cases I, III, IV. 
  	The value of $P_{E}$ is calculated as $P_{E}(t) = \int\!\!\! \int \textbf{P}(x,0,t) \cdot d\textbf{S}_{xz}$, where $\textbf{P}(x,0,t)$ is the Poynting flux vector through the bottom boundary and $d\textbf{S}_{xz}$ represents the area at the bottom boundary. The direction of $d\textbf{S}_{xz}$ is along the $y$-axis. Since all the variables are only functions of x, y in space and they do not change along z-direction, we then assume $d\textbf{S}_{xz}=dxl_z\mathbf{\hat{e_y}}$ and $l_z=100L_0$. From Figure 8, we notice that the energy emerges fastest in Case IV and the corresponding maximum of P is also the largest in three cases. \newline
   
  From Table~1, we notice that the jet lifetime in Case~III is about 40 min which is the longest in all the cases. As shown in Figure~3, the maximum velocity along the jet direction in Case III is 295~km~s${^{-1}}$, which is slower than that in Case I. The highest temperature in Case~III is $1.7$~MK, which is slightly lower than that in Case I. The vortex-like blobs start to form at $t=950$~s in Case~III, about 150~s later than in Case I. Magnetic islands in the main current sheet of Case~III start to appear at $t=1310 $~s, 60~s later than in Case I. We also find that the jet lifetime in Case~IV is about 32 min, it is the shortest one in all the cases. The maximum velocity along the jet direction is $335$~km\,s$^{-1}$, which is a little bit faster than that in Case~I. The maximum temperature in Case~IV is 2.2~MK, which is higher than in Case~I. The vortex-like blobs start to form at $t=750$~s, 50~s earlier than in Case~I, and the magnetic islands appear at  about $t=1190$~s, $60$~s earlier than in Case~I. Figure~3 also  indicates  that the maximum velocity along the jet direction is slower when the flux emerging speed is slower before the K-H instability initiates. \newline
  
 Comparing various features and behavious of magnetic configurations in three cases, we realize that the slower the emerging speed of the magnetic field is, the longer the life time of the jet is, and the later the vortex-like blobs and the magnetic islands in the main current sheet appear. A slower emerging speed yields a lower maximum speed and lower maximum plasma temperature. From Figure 9, we can also see that the smaller vortex-like structures are resulted from slower emerging speed.  \newline

 \section{Conclusion}
   \label{sect:analysis}
   On the basis of the 2D MHD coronal jet simulations by \cite{2017ApJ...841...27N}, we included guide-field  in the $z$-direction in this work to investigate the response  corona to the new emerging flux. Detailed analysis of the K-H instability and comparisons with previous works have been conducted. We have also studied the effect of guide field and the flux emerging speed on the jet formation and the K-H instability. The main conclusions from our numerical simulations are as follow:\newline
   
   1. For the coronal EUV jet with the plasma density in the range from $ 6\times10^{15}$ to $5\times10^{16}$~m$^{-3}$, the maximum values of magnetic field of 15 G and jet speed of 265~km~s${^{-1}}$, the Alfv\'{e}n Mach number reaches 5-14 along the jet direction. The K-H instability can take place in such an EUV coronal jet. Our numerical results confirm the theoretical model and speculating results of \cite{2015ApJ...813..123Z}.\newline
   
   2.The vortex-like blobs are destroyed as a result of the small-scale reconnection processes among the rolled-up magnetic field inside these blobs, as well as the slowing down of the upward reconnection outflow from the main current sheet after magnetic islands appear. A strong guide field changes the reconnection outflow pressure balance structures to prevent an apparent shock structure from being invoked, but helps magnetic islands appear earlier than in the case without guide field or with weak guide field, which further prevents the occurrence of the K-H instability and the vortex-like blob.\newline
   
   3. The speed of the new emerging flux affects the occurrence and development of the K-H instability as well. With the other parameters for the environment being given, the faster the flux emerges, the shorter the lifetime of the jet is, the higher the speed and the maximum temperature of the jet are, the earlier the magnetic island in the main current sheet and the K-H instability occur, and the larger and hotter the vortex-like structures are. \newline

 In this work, guide field is imbedded in the magnetic configuration of interest, and various of several important parameters, as well as the associated behaviors of the corona jet system in 2.5D have been investigated. We expect to perform the true 3D numerical experiments in the future for looking into the formation of the flux rope (the counterpart of the magnetic island in 2D) in the main current sheet, and further studying the K-H instability in the poloidal direction as a result of the rotation of the jet. \newline

 \begin{acknowledgements}
 This work is supported by the NSFC Grants 11573064, 11203069, 11333007, 11303101 and 11403100; Program 973 grant 2013CBA01503; the NSFC-CAS Joint Fund U1631130; the CAS giant QYZDJ-SSW-SLH012; the Western Light of Chinese Academy of Sciences 2014; the Youth Innovation Promotion Association CAS 2017; Key Laboratory of Solar Activity grant KLSA201404. We have used the NIRVANA code v3.6 and v3.8 developed by Udo Ziegler at the Leibniz-Institut f\"ur Astrophysik  Potsdam. The authors also gratefully acknowledge the computing time granted by the Yunnan Astronomical Observatories and the National Supercomputer Center in Guangzhou, the NSFC-Guangdong Joint Fund U1501501 (nsfc2015-460, nsfc2015-463). Some calculations of this work were performed on the Tianhe-1 supercomputer of the National Supercomputer Center in Tianjin.
 \end{acknowledgements}

  \end{spacing}

   \label{lastpage}
   
   \newpage
   \begin{table}
   	\begin{center}
   	        \caption[]{ The important parameters in different cases in our simulations. $b_0$ is the initial background magnetic field, $t_\mathrm{life}$ the lifetime of the jet, $v_\mathrm{||max}$ is the maximum velocity along the jet direction, $t_\mathrm{v}$ is the time when the vortex structure starts to appear, $t_\mathrm{i}$ is the time when the magnetic  island starts to appear, $N$ is the number of vortex structures, $T_\mathrm{max}$ is the maximum temperature in the jet.}\label{Tab:publ-works}

   		\begin{tabular}{clclclclcl}
   			\hline\noalign{\smallskip}
  			Case &  $t_1\mathrm{(s) }$     & ${b_z}$ &$t_{\mathrm{life}}\mathrm{(min)}$&$v_{||max}\mathrm{(km/s)}$&$t_v(\mathrm{s})$&$N$&$t_i(\mathrm{s})$&$T_{max}(\mathrm{MK})$                     \\
  			\hline\noalign{\smallskip}
   			I  &500 & $0.05b_0 $    & 37 &320&800&2&1250&1.8   \\ % new variable
   			II  & 500  & $b_0$&33&275&850&2&1080&1.6                  \\
   			III  & 700  &   $0.05b_0 $ & 40&295&950&3&1310&1.7                \\
  			IV &350 &$0.05b_0$ &32&335 &750 &3 &1190 & 2.2       \\
   			\hline\noalign{\smallskip}
	       \end{tabular}
   	\end{center}
   \end{table}
  
  \begin{table}
  	\begin{center}
  		\caption[]{ Comparies of values of some important parameters studied by Zaqrarshvili et al (2015) and by us in Case I. Our results are selected at $t=784.3$~s in the simulation just before the K-H instability takes place. Both the jet speed and the Alfv\'{e}n speed are along the jet direction.}\label{Tab:publ-works}

  		\begin{tabular}{clclclclcl}
  			\hline\noalign{\smallskip}
  			Parameters &  Zaqarashvili's     & Case I (average) &  CaseI (maximum)                    \\
  			\hline\noalign{\smallskip}
  			Magnetic field(G)  &10 & 5    & 15 &  \\ % new variable
  			Alfv\'{e}n speed(km/s)  & 220   &50  & 250          \\
  			Jet speed (km/s) & 250   &  160  & 265               \\
  			Density (m$^{-3}$ )&$10^{16}-10^{17}$
  			&$2.8\times10^{16}$
  			& $   5\times 10^{16}$   \\
  			\hline\noalign{\smallskip}
  		\end{tabular}
  	\end{center}
  \end{table}
  
   \begin{figure}
   	\centering
   	\includegraphics[width=\textwidth, angle=0]{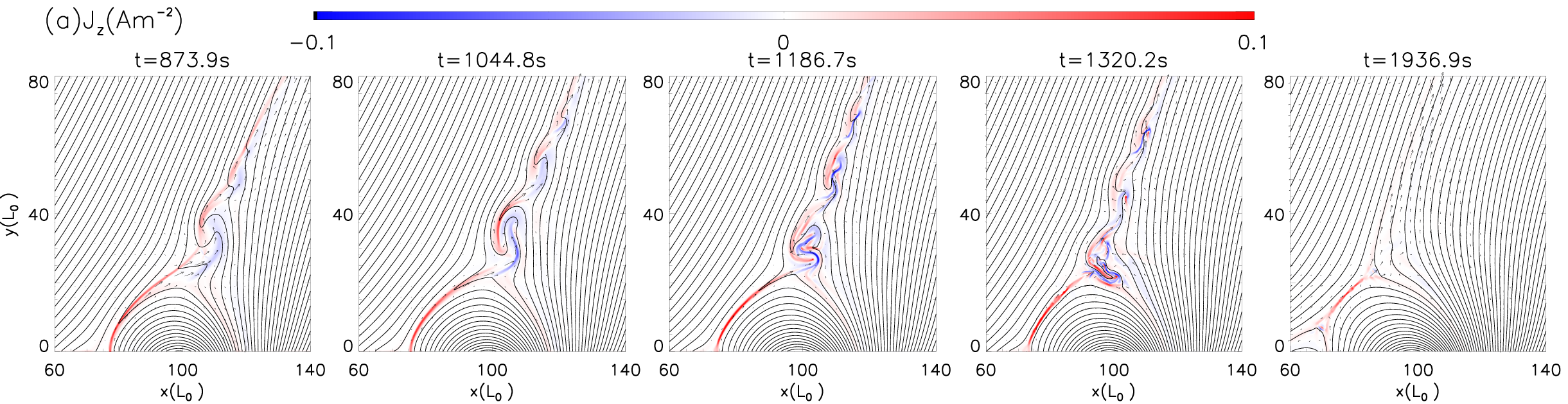}
   	\includegraphics[width=\textwidth, angle=0]{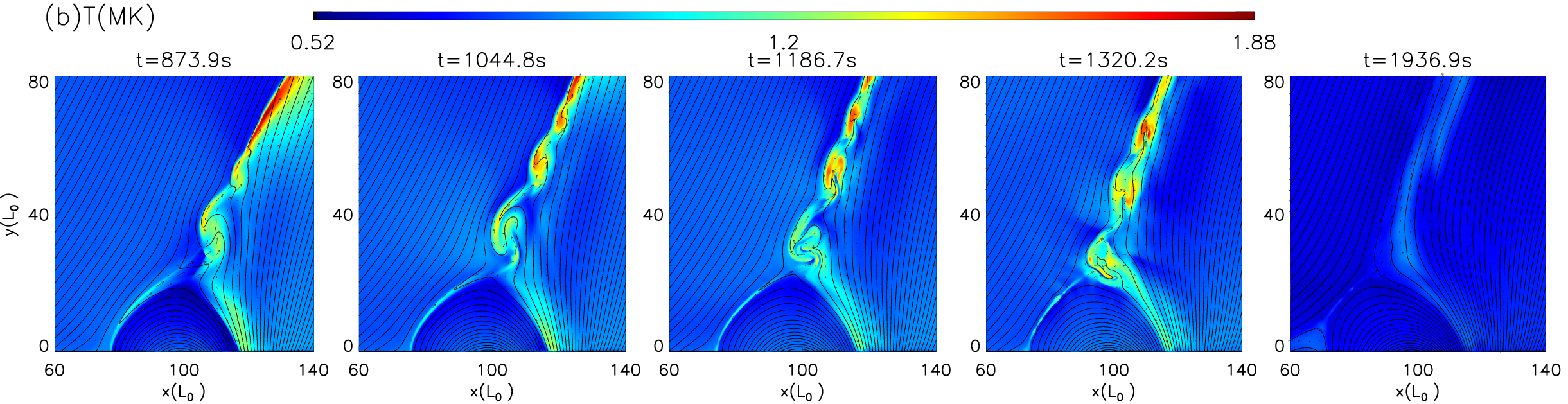}
   	\includegraphics[width=\textwidth, angle=0]{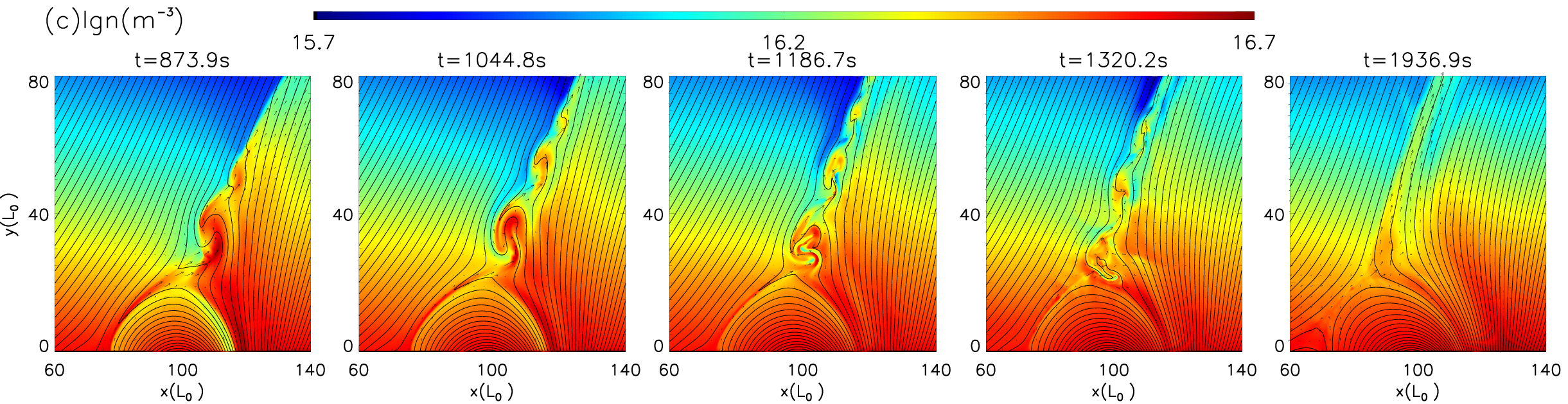}
   	\includegraphics[width=\textwidth, angle=0]{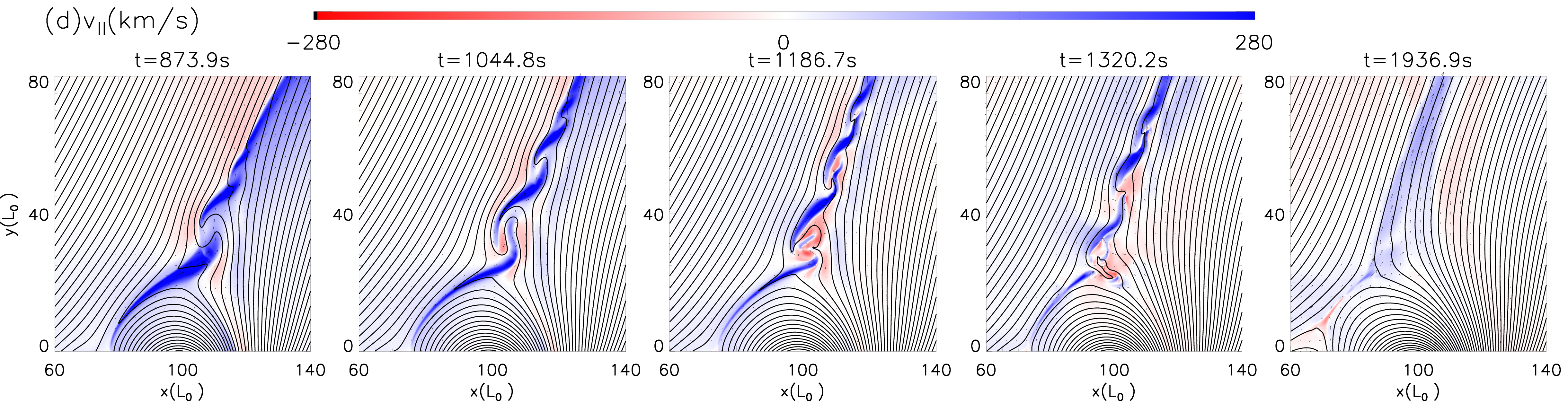}
   	 	\includegraphics[width=\textwidth, angle=0]{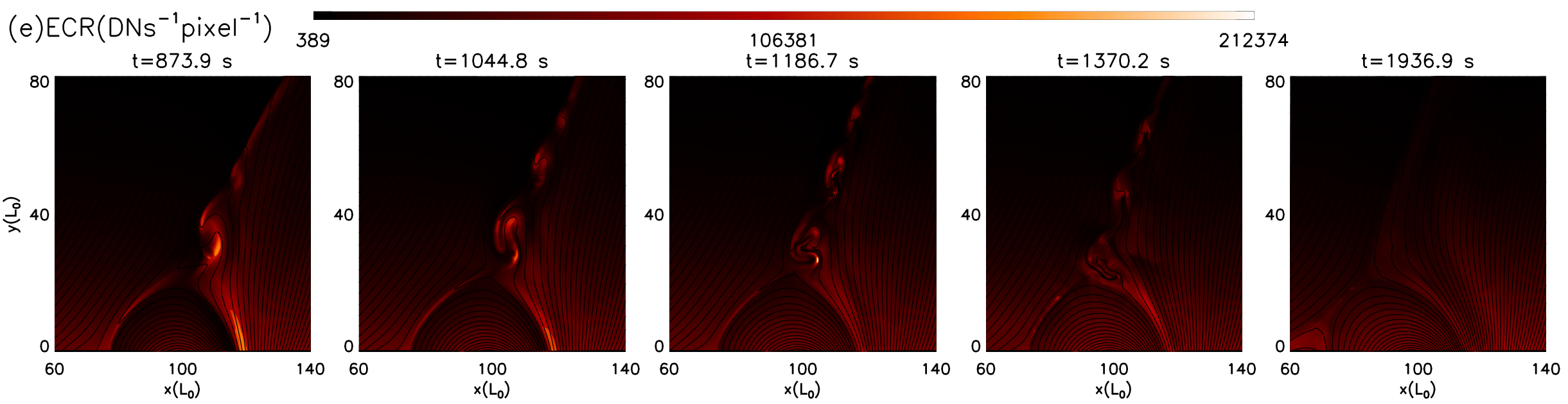}
   	\caption{Distributions of different variables at five different times in Case I, (a) current density, $J_z$, (b) temperature,T, (c) logarithm of plasma number density, $\lg n$, (d) velocity along the jet direction, $v_{||max}$. (e) The distributions of the emission count rate in the AIA 211\,\AA\, channel at five different times in Case~I\@. Continvous black curves represent the magnetic fields and the black arrows represent the velocity vector in each panels.}
   	
  	\label{Fig1}
   \end{figure}

   \begin{figure}
   	\centering
   	\includegraphics[width=\textwidth, angle=0]{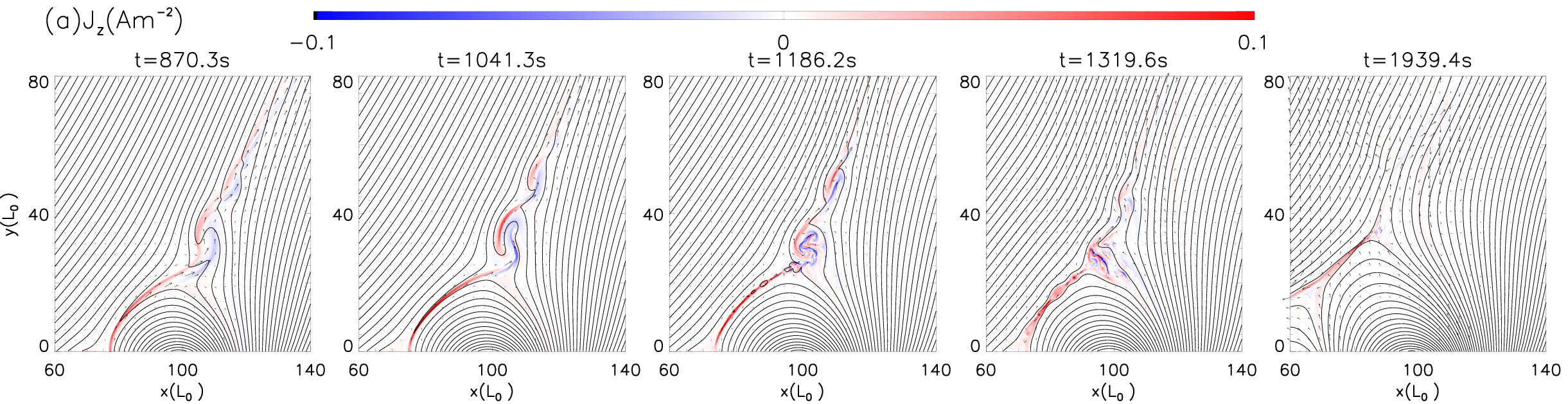}
   	\includegraphics[width=\textwidth, angle=0]{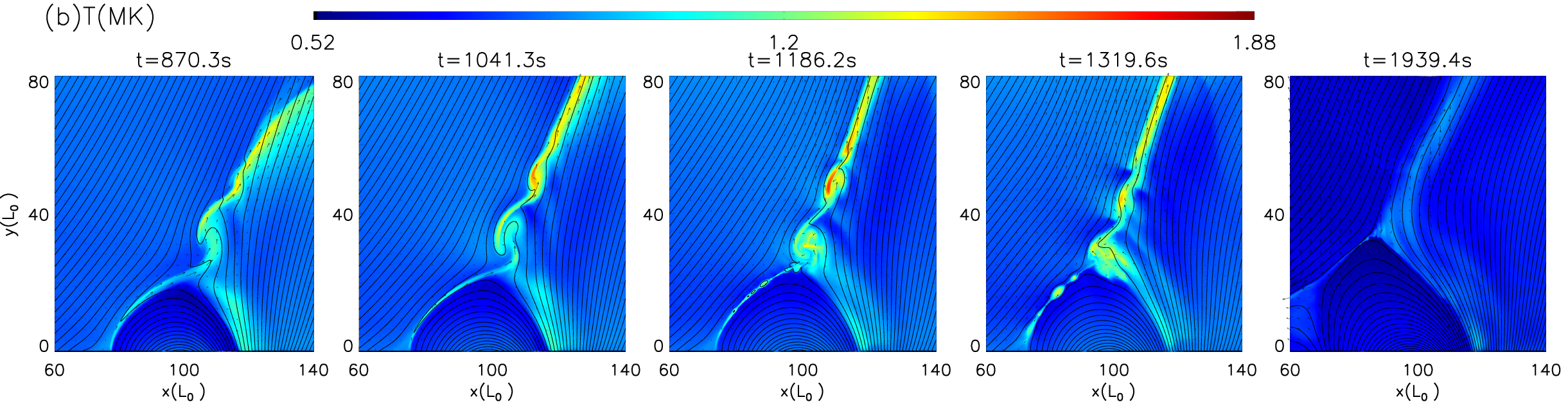}
   	\includegraphics[width=\textwidth, angle=0]{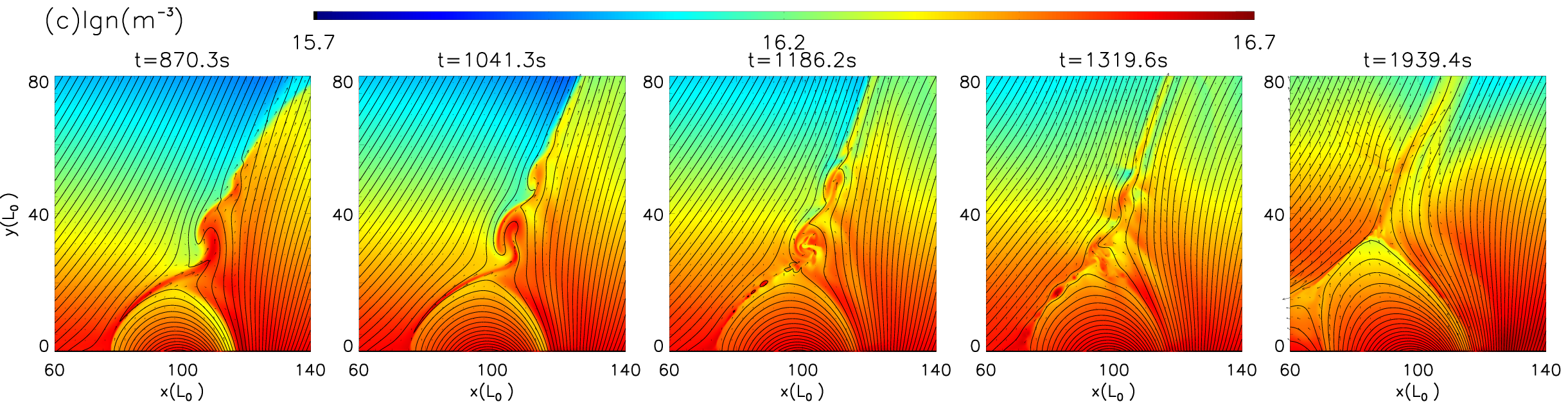}
  	\includegraphics[width=\textwidth, angle=0]{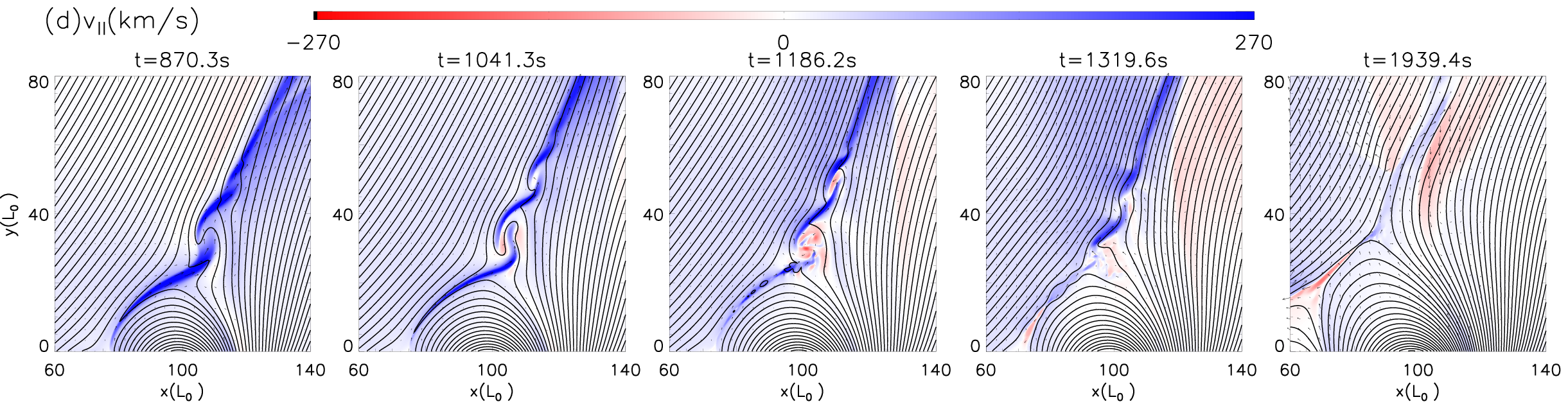}
   	 	\includegraphics[width=\textwidth, angle=0]{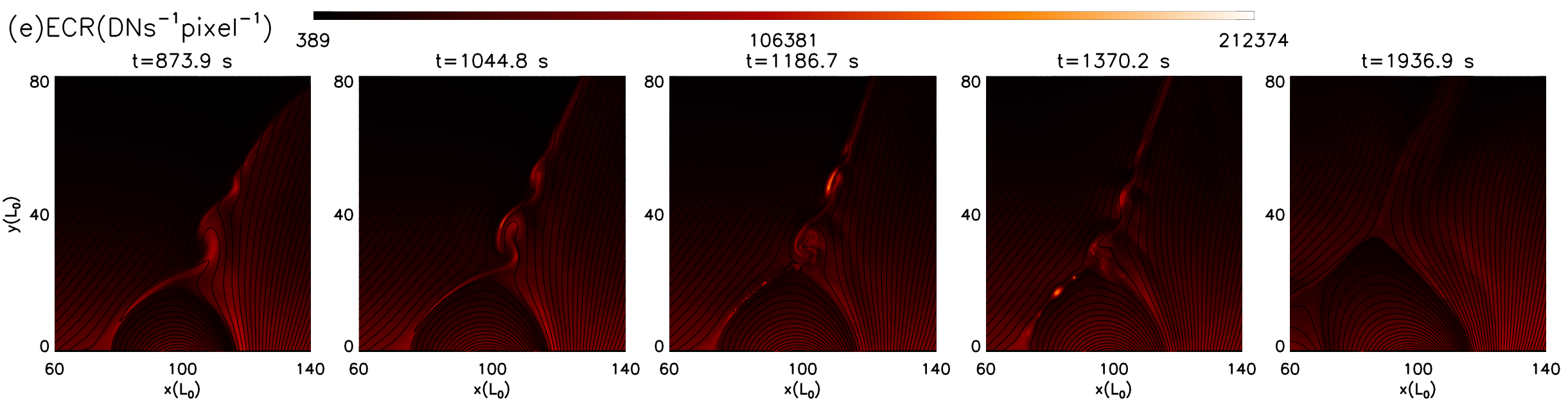}
   	\caption{Same as Figure 1 for Case II. }
   	\label{Fig2}
   \end{figure}
   
    \begin{figure}
   	\centering
   	\includegraphics[width=0.7\textwidth, angle=0]{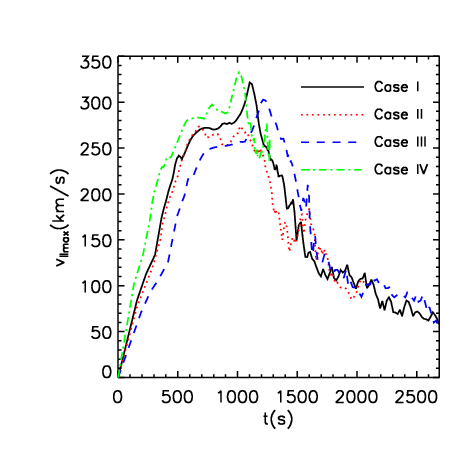}
   	\caption{The maximum velocities along the jet direction versus time for four cases. }
   	\label{Fig3}
   \end{figure}
   
   \begin{figure}
   	\centering
   	\includegraphics[width=\textwidth, angle=0]{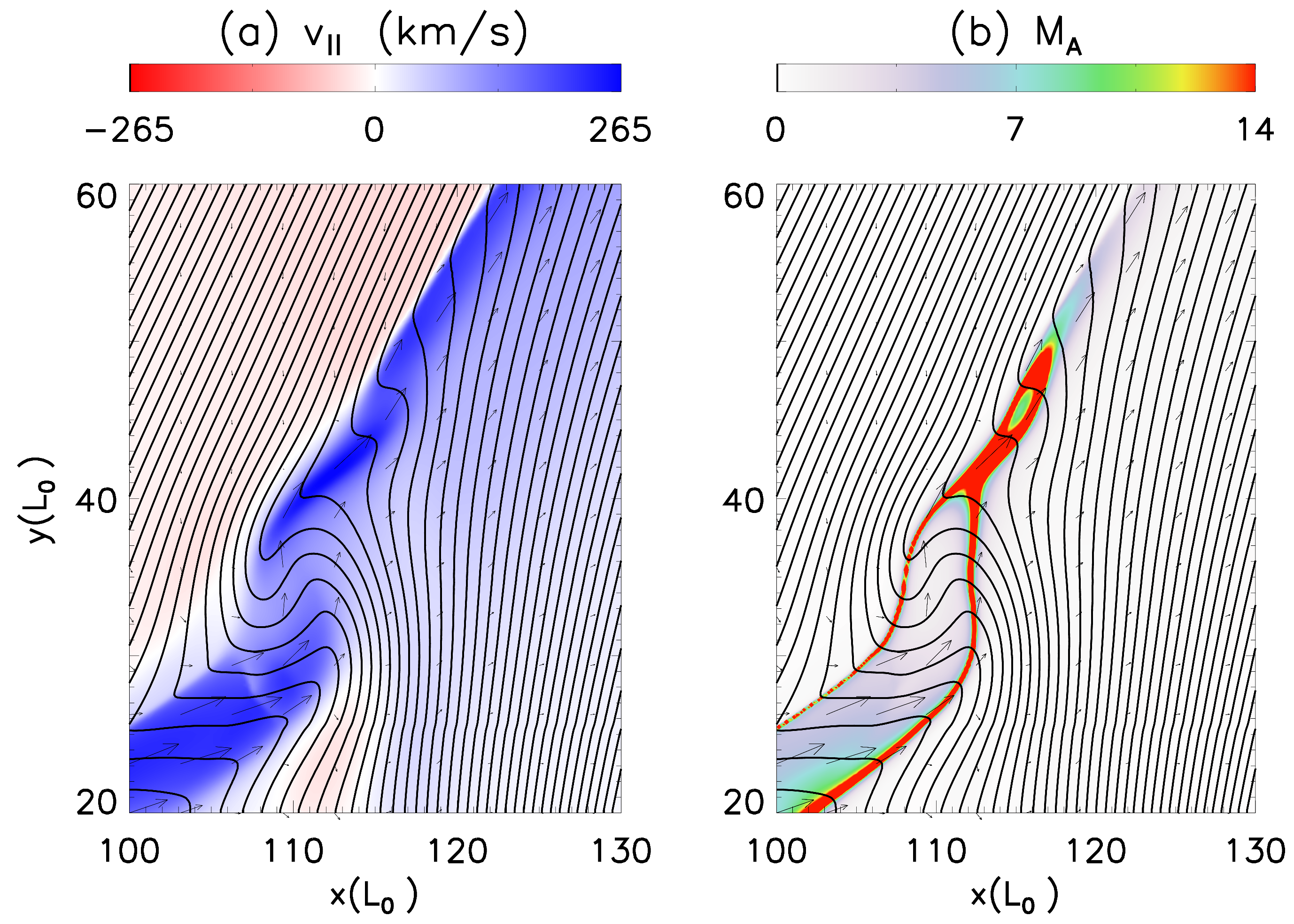}
   	\caption{Distributions of the velocity and $M_A$ along the jet direction for Case I at $t=784.3$~s. }
   	\label{Fig4}
   \end{figure}
   
  \begin{figure}
   	\centering
  	\includegraphics[width=\textwidth, angle=0]{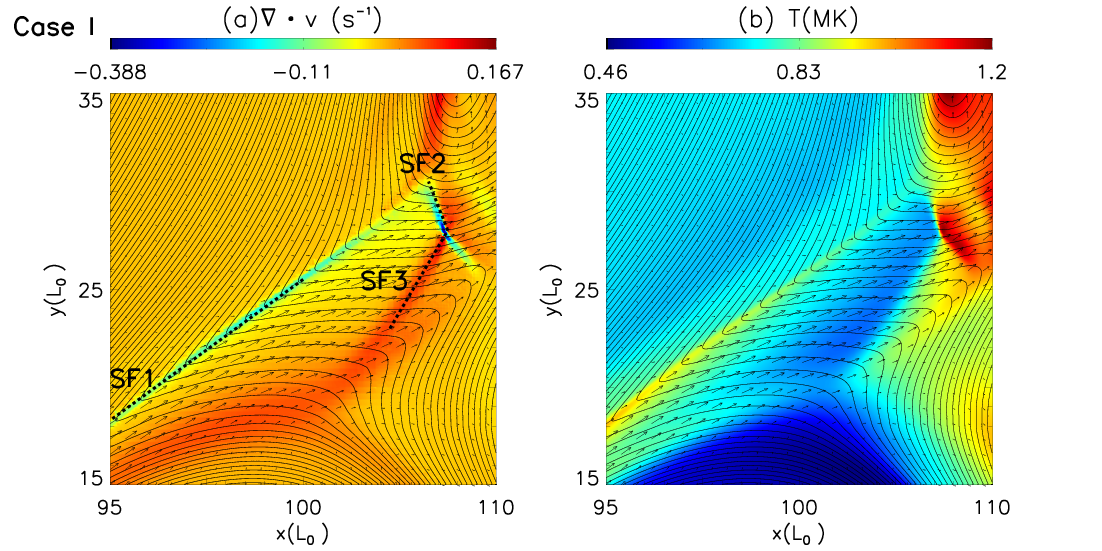}
  	\includegraphics[width=\textwidth, angle=0]{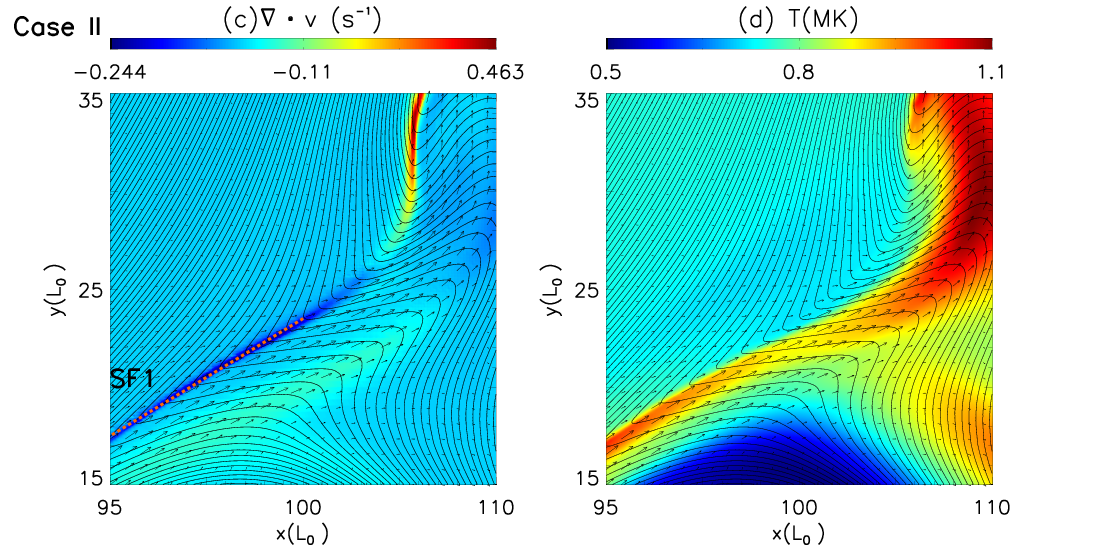}
   	\caption{Distributions of the velocity divergence (a) and temperature (b) in the reconnection outflow region of the main current sheet at $t=801.7 $~s in Case~I ; the same for Case II in panels (c) and (d) at $t=794.8$~s. }
   	\label{Fig5}
   \end{figure}
   
     \begin{figure}
   	\centering
   	\includegraphics[width=\textwidth,
   	angle=0]{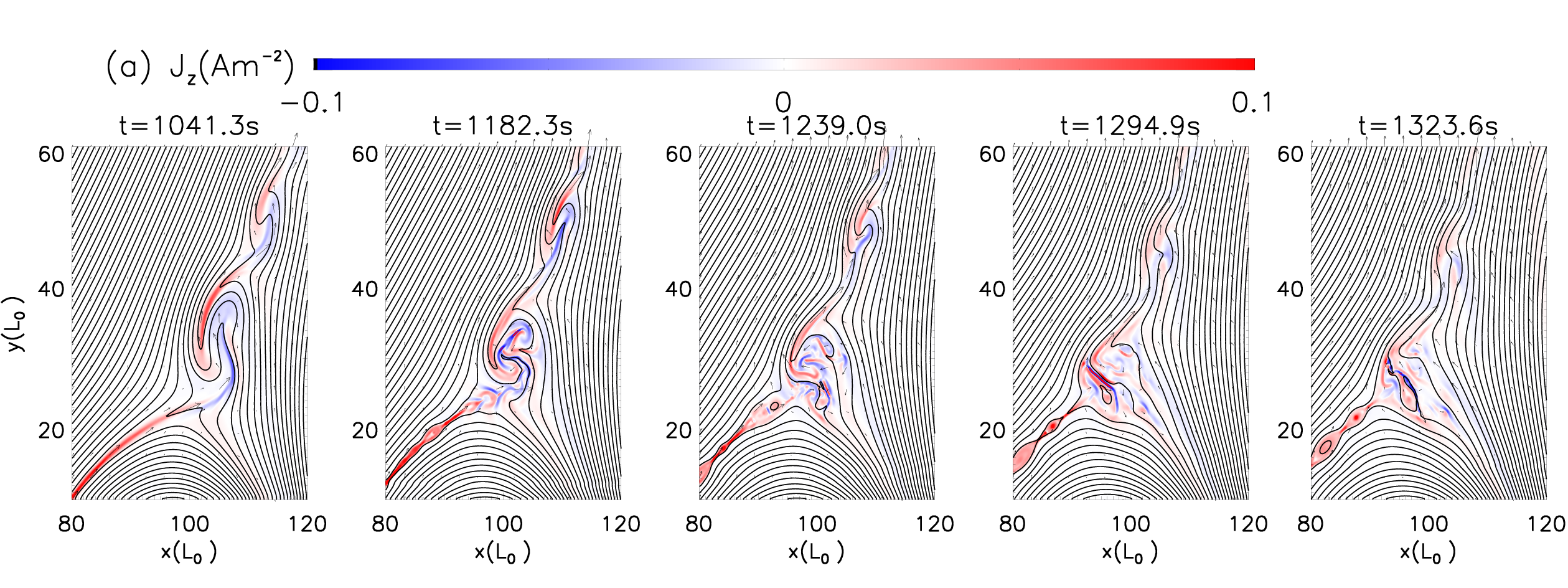}
   	\includegraphics[width=\textwidth, angle=0]{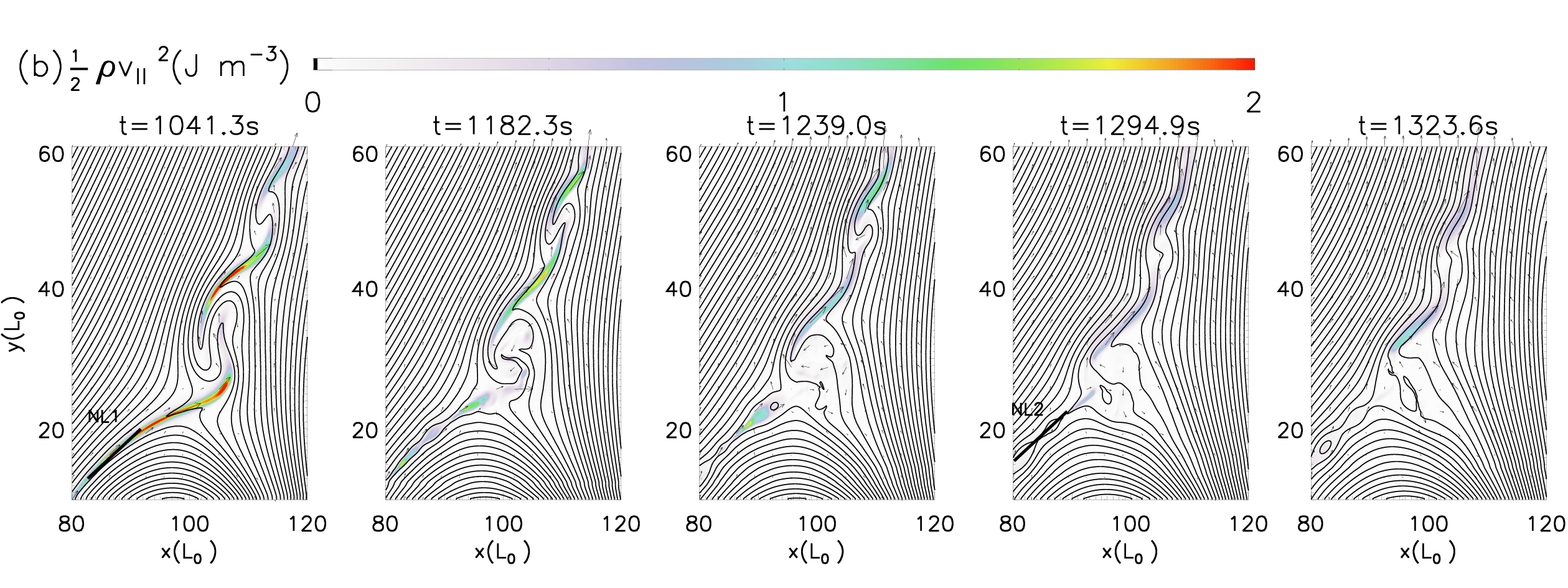}
      \begin{minipage}[c]{0.5\textwidth}
   	\centering
   	\includegraphics[height=7cm,width=7cm]{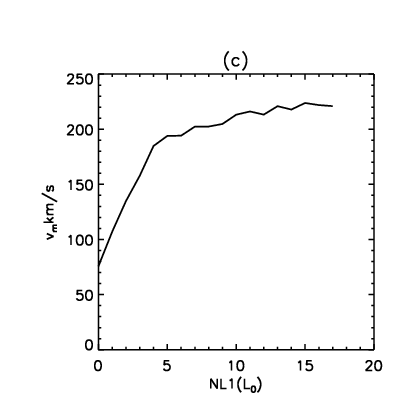}
  \end{minipage}%
  \begin{minipage}[c]{0.5\textwidth}
   \centering
  	\includegraphics[height=7cm,width=7cm]{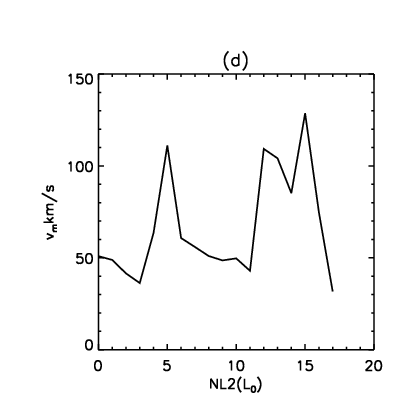}
  \end{minipage}
  
   	\caption{(a) Distributions the current density, $J_z$,  (b) the kinetic energy ${\rho {v_{||}}^2}/{2}$ along the jet direction before and after the magnetic island in the main current sheet appears for Case~II, (c) velocity along the current sheet direction NL1at $t=1041.3$~s,(d) velocity along the current sheet direction NL2 at $t=1294.9$~s. }
   	\label{Fig6}
   \end{figure} 
   
     \begin{figure}
   	\centering
   	\includegraphics[width=\textwidth, angle=0]{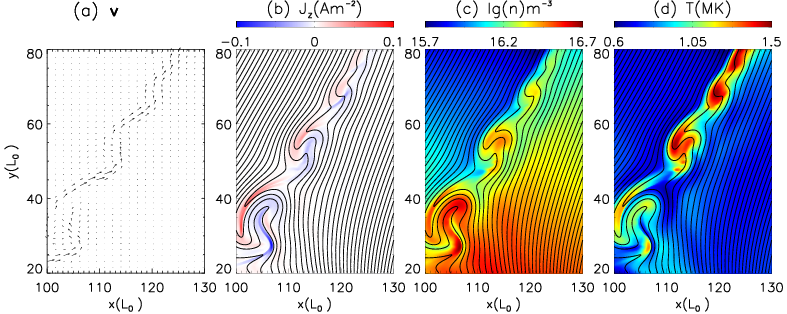}
   	\includegraphics[width=\textwidth, angle=0]{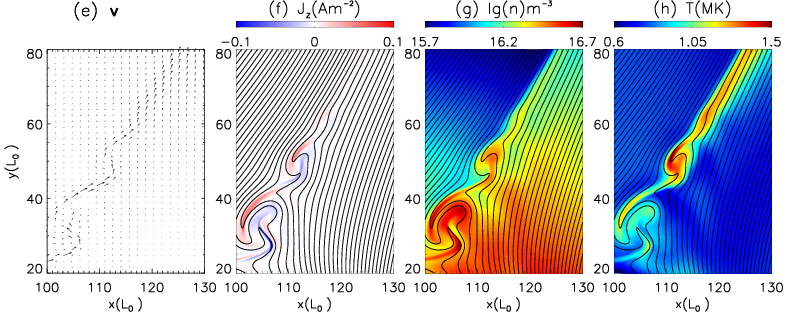}
   	\caption{The distribution of velocity vector,current density $J_z$, logarithm of plasma number density $\lg n$, temperature $T$. Panels (a), (b), (c), and (d) are for Case I at $t=1082$ s; and panels (e), (f), (g), and (h) are for Case II at $t=1079$ s.}
  	\label{Fig7}
   \end{figure}

       \begin{figure}
   	\centering
   	\includegraphics[width=0.7\textwidth, angle=0]{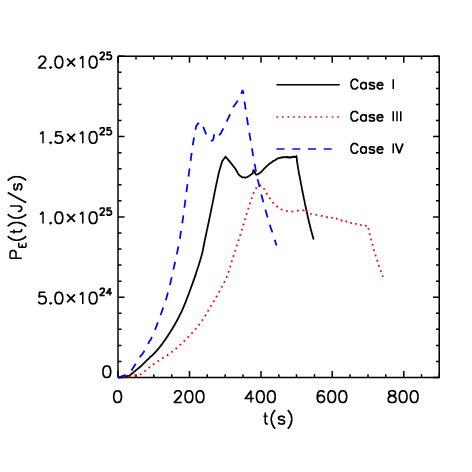}
   	\caption{Variations of the electromagnetic energy emerging through the bottom boundary versus time in Cases I, III, IV.}
   	\label{Fig8}
   \end{figure}

  %\B{(C) velocity along the arrow NL1,(d) velocity along the arrow NL2.
   
   \begin{figure}
   	\centering
   	\includegraphics[width=\textwidth, angle=0]{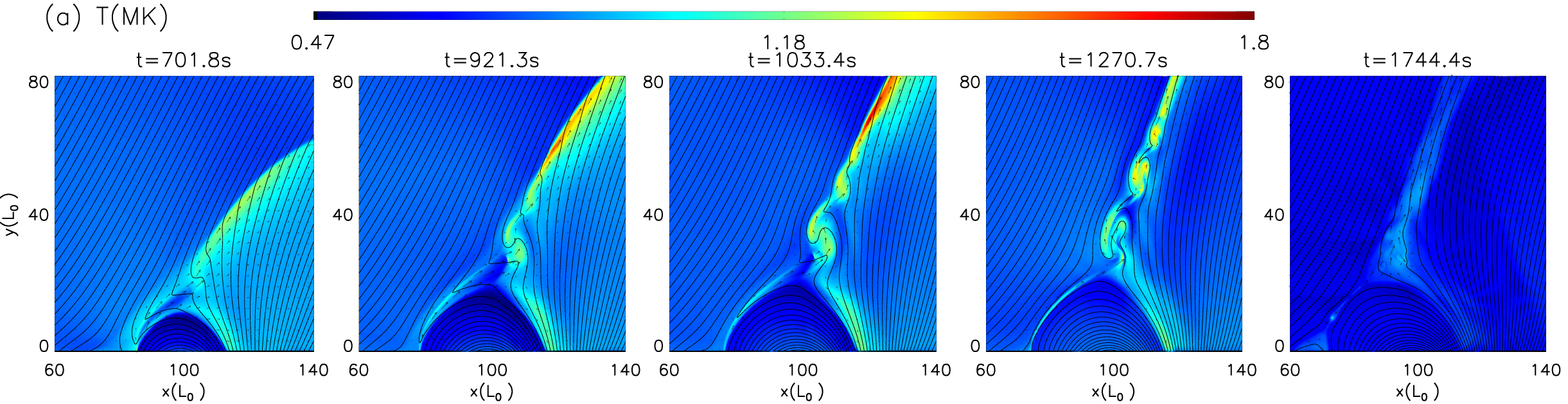}
   	\includegraphics[width=\textwidth, angle=0]{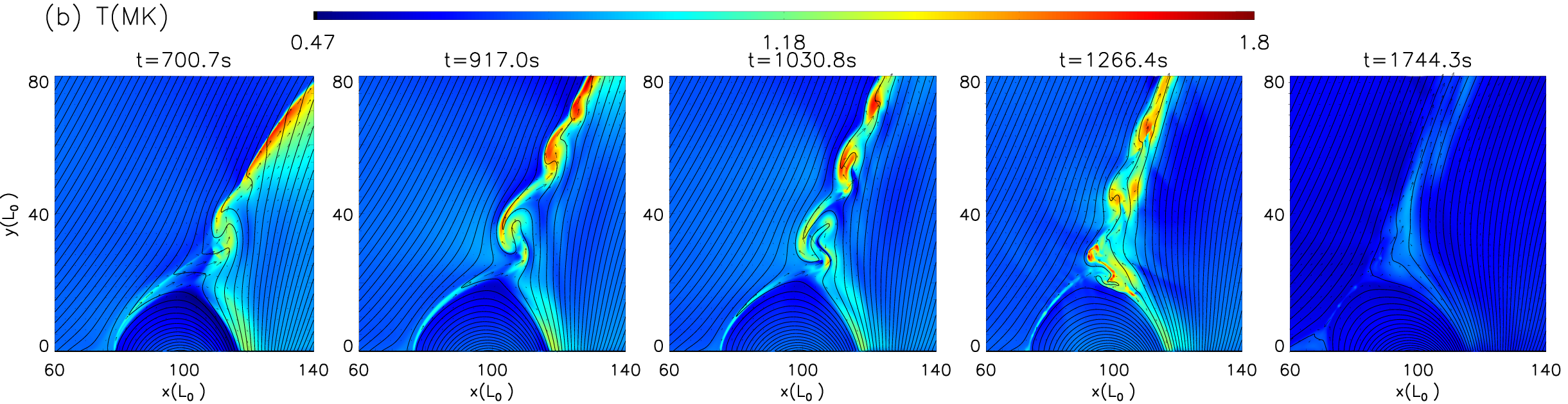}
   	\caption{Distributions of the temperature $T$ at five different times for Case~III (upper row) and Case~IV (bottom row).}
   	\label{Fig9}
   \end{figure}
\label{lastpage}


\begin{thebibliography}{99}
   
   	
     \bibitem[Alexander \& Fletcher(1999)]{1999SoPh..190..167A} Alexander, D., \& Fletcher, L.\ 1999, \solphys, 190, 167 
	
	\bibitem[B{\'a}rta et al.(2011)]{2011ApJ...737...24B} B{\'a}rta, M., B{\"u}chner, J., Karlick{\'y}, M., \& Sk{\'a}la, J.\ 2011, \apj, 737, 24 
	
	\bibitem[Bhattacharjee et al.(2009)]{2009PhPl...16k2102B} Bhattacharjee, A., Huang, Y.-M., Yang, H., \& Rogers, B.\ 2009, Physics of Plasmas, 16, 112102
   	
   	
   	\bibitem[Chen \& Shibata(2000)]{2000ApJ...545..524C} Chen, P.~F., \& Shibata, K.\ 2000, \apj, 545, 524 
   
   	\bibitem[Comisso \& Bhattacharjee(2016)]{2016JPlPh..82f5901C} Comisso, L., \& Bhattacharjee, A.\ 2016, Journal of Plasma Physics, 82, 595820601 
   	
   	\bibitem[Ding et al.(2010)]{2010A&A...510A.111D} Ding, J.~Y., Madjarska, M.~S., Doyle, J.~G., \& Lu, Q.~M.\ 2010, \aap, 510, A111 
   	
   	
   	\bibitem[Forbes \& Priest(1984)]{1984SoPh...94..315F} Forbes, T.~G., \& Priest, E.~R.\ 1984, \solphys, 94, 315 
   	
   	
   	\bibitem[Foullon et al.(2013)]{2013ApJ...767..170F} Foullon, C., Verwichte, E., Nykyri, K., Aschwanden, M.~J., \& Hannah, I.~G.\ 2013, \apj, 767, 170 
   	
   	
	\bibitem[Innes et al.(2015)]{2015ApJ...813...86I} Innes, D.~E., Guo, L.-J., Huang, Y.-M., \& Bhattacharjee, A.\ 2015, \apj, 813, 86 
	
   	
   	\bibitem[Jeong et al.(2000)]{2000ApJ...529..536J} Jeong, H., Ryu, D., Jones, T.~W., \& Frank, A.\ 2000, \apj, 529, 536 
	
	\bibitem[Jiang et al.(2012)]{2012ApJ...751..152J} Jiang, R.-L., Fang, C., \& Chen, P.-F.\ 2012, \apj, 751, 152
   	
   	
   	\bibitem[Jones et al.(1997)]{1997ApJ...482..230J} Jones, T.~W., Gaalaas, J.~B., Ryu, D., \& Frank, A.\ 1997, \apj, 482, 230 
   	
   	
   	\bibitem[Keppens et al.(1999)]{1999JPlPh..61....1K} Keppens, R., T{\'o}th, G., Westermann, R.~H.~J., \& Goedbloed, J.~P.\ 1999, Journal of Plasma Physics, 61, 1 
   	
   	\bibitem[Kuridze et al.(2016)]{2016ApJ...830..133K} Kuridze, D., Zaqarashvili, T.~V., Henriques, V., et al.\ 2016, \apj, 830, 133 
   	
 %%Lord Kelvin (William Thomson) 1871, Phil. Mag. 42, 362
     \bibitem[Li et al.(2017)]{2017ApJ...842L..20L} Li, H., Jiang, Y., Yang, J., et al.\ 2017, \apjl, 842, L20
     
   \bibitem[Mei et al.(2012)]{2012MNRAS.425.2824M} Mei, Z., Shen, C., Wu, N., et al.\ 2012, \mnras, 425, 2824 
    
     
   	\bibitem[Moore et al.(2010)]{2010ApJ...720..757M} Moore, R.~L., Cirtain, J.~W., Sterling, A.~C., \& Falconer, D.~A.\ 2010, \apj, 720, 757 
	
	\bibitem[Moreno-Insertis \& Galsgaard(2013)]{2013ApJ...771...20M} Moreno-Insertis, F., \& Galsgaard, K.\ 2013, \apj, 771, 20 
   	
   	\bibitem[Nemati et al.(2017)]{2017ApJ...835..191N} Nemati, M.~J., Wang, Z.-X., \& Wei, L.\ 2017, \apj, 835, 191
	
	
	
   	\bibitem[N{\'o}brega-Siverio et al.(2016)]{2016ApJ...822...18N} N{\'o}brega-Siverio, D., Moreno-Insertis, F., \& Mart{\'{\i}}nez-Sykora, J.\ 2016, \apj, 822, 18 
   	
	\bibitem[Ni et al.(2013)]{2013PhPl...20f1206N} Ni, L., Lin, J., \& Murphy, N.~A.\ 2013, Physics of Plasmas, 20, 061206
	
	\bibitem[Ni et al.(2015)]{2015ApJ...799...79N} Ni, L., Kliem, B., Lin, J., \& Wu, N.\ 2015, \apj, 799, 79
   	
          \bibitem[Ni et al.(2017)]{2017ApJ...841...27N} Ni, L., Zhang, Q.-M., Murphy, N.~A., \& Lin, J.\ 2017, \apj, 841, 27 
   	
   	
   	\bibitem[Ofman \& Thompson(2011)]{2011ApJ...734L..11O} Ofman, L., \& Thompson, B.~J.\ 2011, \apjl, 734, L11 
   	
   	\bibitem[Priest(2014)]{2014masu.book.....P} Priest, E.\ 2014, Magnetohydrodynamics of the Sun, by Eric Priest, Cambridge, UK: Cambridge University Press, 2014, p177-188
   
   	\bibitem[Raouafi et al.(2016)]{2016SSRv..201....1R} Raouafi, N.~E., Patsourakos, S., Pariat, E., et al.\ 2016, \ssr, 201, 1 
   	
   	
   	\bibitem[Roy \& Tang(1975)]{1975SoPh...42..425R} Roy, J.-R., \& Tang, F.\ 1975, \solphys, 42, 425 
	
	\bibitem[Shen et al.(2011)]{2011ApJ...735L..43S} Shen, Y., Liu, Y., Su, J., \& Ibrahim, A.\ 2011, \apjl, 735, L43
	
	\bibitem[Shen et al.(2011)]{2011ApJ...737...14S} Shen, C., Lin, J., \& Murphy, N.~A.\ 2011, \apj, 737, 14 
	
	\bibitem[Shen et al.(2017)]{2017ApJ...851...67 } Shen, Y., Liu, Y., Tian, Z., \& Qu, Z.\ 2017, \apj, 851, 67 
   	
   	\bibitem[Shibata \& Murdin(2000)]{2000eaa..bookE2272S} Shibata, K., \& Murdin, P.\ 2000, Encyclopedia of Astronomy and Astrophysics,  
   	
   	
   	\bibitem[Shibata et al.(1992)]{1992PASJ...44L.173S} Shibata, K., Ishido, Y., Acton, L.~W., et al.\ 1992, \pasj, 44, L173 
   	
   	
   	\bibitem[Shibata et al.(2007)]{2007Sci...318.1591S} Shibata, K., Nakamura, T., Matsumoto, T., et al.\ 2007, Science, 318, 1591 
   	
   	\bibitem[Spitzer (1962)]{Spitzer62} Spitzer, L.,~Jr. 1962, Physics of Fully Ionized Gases (New York: Interscience)
   	
   	\bibitem[Tian \& Chen(2016)]{2016ApJ...824...60T} Tian, C., \& Chen, Y.\ 2016, \apj, 824, 60 
	
	 \bibitem[Wyper et al.(2016)]{2016ApJ...827....4W} Wyper, P.~F., DeVore, C.~R., Karpen, J.~T., \& Lynch, B.~J.\ 2016, \apj, 827, 4 
	
	\bibitem[Yang et al.(2013)]{2013ApJ...777...16Y} Yang, L., He, J., Peter, H., et al.\ 2013, \apj, 777, 16
   	
   	\bibitem[Zaqarashvili et al.(2015)]{2015ApJ...813..123Z} Zaqarashvili, T.~V., Zhelyazkov, I., \& Ofman, L.\ 2015, \apj, 813, 123 
   	\bibitem[Zhang \& Ji(2014)]{2014A&A...567A..11Z} Zhang, Q.~M., \& Ji, H.~S.\ 2014, \aap, 567, A11
   	
   	\bibitem[Zhang et al.(2016)]{2016SoPh..291..859Z} Zhang, Q.~M., Ji, H.~S., \& Su, Y.~N.\ 2016, \solphys, 291, 859 
   	
   	
   	\bibitem[Zhang \& Zhang(2017)]{2017ApJ...834...79Z} Zhang, Y., \& Zhang, J.\ 2017, \apj, 834, 79 
   	
   	
   	\bibitem[Ziegler(2011)]{2011JCoPh.230.1035Z} Ziegler, U.\ 2011, Journal of Computational Physics, 230, 1035 
   	
   	
   	\bibitem[Ziegler(2008)]{2008CoPhC.179..227Z} Ziegler, U.\ 2008, Computer Physics Communications, 179, 227 
   	
   	
   \end{thebibliography}
\end{document}